\documentclass[10pt,twocolumn,letterpaper]{article}
\usepackage[pagenumbers]{wacv} 

\usepackage{latexsym}
\usepackage{amssymb}
\usepackage{amsmath}
\usepackage{amsthm}
\usepackage{booktabs}
\usepackage{enumitem}
\usepackage{graphicx}
\usepackage{color}
\usepackage[ruled, vlined]{algorithm2e}
\usepackage{multirow}
\usepackage{threeparttable}
\usepackage{graphicx}
\usepackage{amsmath}
\usepackage{amssymb}
\usepackage{booktabs}
\usepackage{morefloats}
\extrafloats{100}
\usepackage{etex}
\usepackage{algorithmic}
\usepackage[pagebackref,breaklinks,colorlinks]{hyperref}

\usepackage[capitalize]{cleveref}
\crefname{section}{Sec.}{Secs.}
\Crefname{section}{Section}{Sections}
\Crefname{table}{Table}{Tables}
\crefname{table}{Tab.}{Tabs.}

\def\wacvPaperID{2330} 
\def\confName{WACV}
\def\confYear{2025}

\begin{document}

\title{Ancestral Mamba: Enhancing Selective Discriminant Space Model with Online Visual Prototype Learning for Efficient and Robust Discriminant Approach}

\author{Jiahao Qin\\
Xi'an Jiaotong-Liverpool University\\
Suzhou, China\\
{\tt\small Jiahao.qin19@gmail.com}
\and
Feng Liu\\
Shanghai Jiao Tong University\\
Shanghai, China\\
{\tt\small lsttoy@163.com}
\and
Lu Zong\\
Xi'an Jiaotong-Liverpool University\\
Suzhou, China\\
{\tt\small lu.zong@xjtlu.edu.cn}
}
\maketitle

\begin{abstract}
   In the realm of computer graphics, the ability to learn continuously from non-stationary data streams while adapting to new visual patterns and mitigating catastrophic forgetting is of paramount importance. Existing approaches often struggle to capture and represent the essential characteristics of evolving visual concepts, hindering their applicability to dynamic graphics tasks. In this paper, we propose Ancestral Mamba, a novel approach that integrates online prototype learning into a selective discriminant space model for efficient and robust online continual learning. The key components of our approach include Ancestral Prototype Adaptation (APA), which continuously refines and builds upon learned visual prototypes, and Mamba Feedback (MF), which provides targeted feedback to adapt to challenging visual patterns. APA enables the model to continuously adapt its prototypes, building upon ancestral knowledge to tackle new challenges, while MF acts as a targeted feedback mechanism, focusing on challenging classes and refining their representations. Extensive experiments on graphics-oriented datasets, such as CIFAR-10 and CIFAR-100, demonstrate the superior performance of Ancestral Mamba compared to state-of-the-art baselines, achieving significant improvements in accuracy and forgetting mitigation.
\end{abstract}

\section{Introduction}
\begin{figure*}[ht]
\centering
\includegraphics[width=0.8\textwidth]{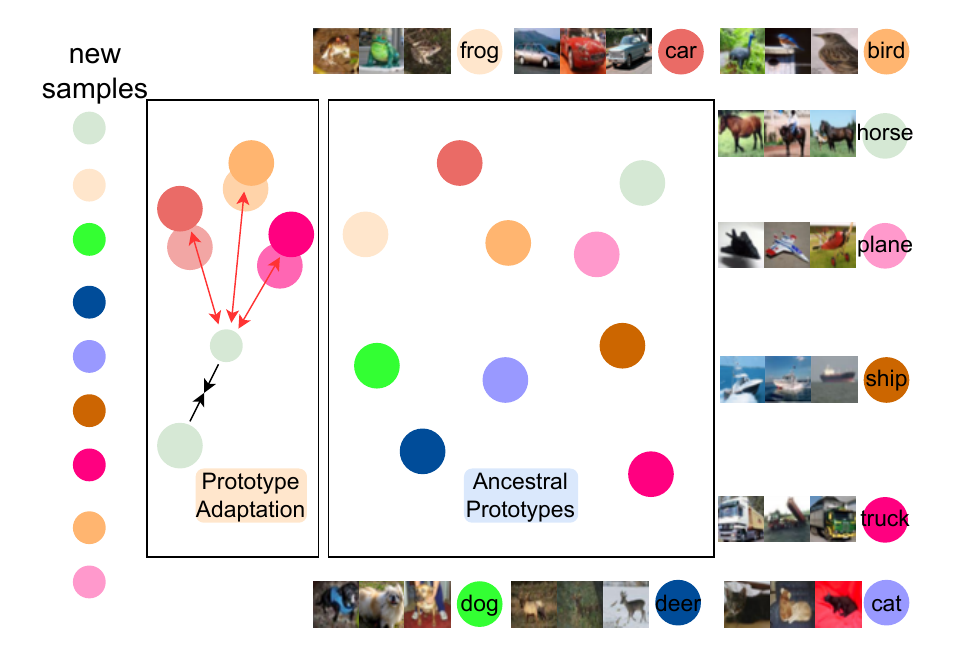}
\caption{The APA module can reduce the distance between the new sample and the most similar prototype and increase the distance between the other prototypes during the learning process.}
\label{fig:fig1}
\end{figure*}

Online continual learning (OCL) aims to learn continuously from a non-stationary data stream while adapting to new data and mitigating catastrophic forgetting \cite{rebuffi2017icarl,buzzega_dark_2020,wang2024comprehensive,liu2023online}. Recently, online prototype Learning (OnPro) \cite{wei2024online} has attracted a lot of attention with its brilliant performance in the OCL field. This paradigm holds immense potential for real-world applications, particularly in the realm of computer graphics, where the ability to process and adapt to evolving visual patterns, shapes, and colours is of paramount importance.



Catastrophic forgetting \cite{han2024adaptive,yu2024adaptive,xiao2023online,wei2024online} stands as a major hurdle in online continual learning, akin to a visual artist abruptly losing previously acquired skills when adapting to new styles. The sudden erosion of knowledge hinders the model's ability to maintain a comprehensive understanding of the visual world. Moreover, the capacity to capture and represent essential characteristics of each class or visual concept is crucial for generalization and robustness, yet current methods often struggle in this regard \cite{mai2021supervised,wei2024online}. Adapting to evolving data distributions is another challenge, as the model needs to continuously update its knowledge without access to previous data \cite{geirhos2020shortcut,gu2020hippo}. Addressing these challenges is of paramount importance for the advancement of online continual learning and its application in real-world scenarios. 

Existing continual learning approaches, such as iCaRL~\cite{rebuffi2017icarl} and ASER~\cite{shim_online_2021}, have made significant strides in addressing these challenges. However, they often grapple with the trade-off between stability and plasticity, struggling to find an optimal balance between preserving previously acquired knowledge and integrating new information. Consequently, there is a pressing need for innovative solutions that can effectively learn discriminative features, efficiently manage limited resources, and seamlessly adapt to evolving data distributions.

To address these limitations, there has been growing interest in developing more efficient and effective online continual learning techniques. One promising direction is the use of prototype-based methods \cite{de2021continual,guo_online_2022,mambaspike_qin}, which have demonstrated strong performance and efficient memory usage in various domains. In particular, the CoPE architecture \cite{de2021continual} introduces a continual prototype evolution approach that learns prototypical class representations in a shared latent space. However, like many online continual learning approaches, CoPE may still be susceptible to shortcut learning \cite{geirhos2020shortcut}, where the model relies on superficial cues rather than learning meaningful representations. This can lead to biased and non-generalizable features, hindering the model's performance on new tasks or under distribution shifts. OnPro \cite{wei2024online} has emerged as a promising framework to address shortcut learning and catastrophic forgetting in online continual learning. OnPro encourages the model to learn features that are close to the corresponding prototypes and far from others; OnPro promotes the learning of discriminative and generalizable representations. 


\begin{figure*}[ht]
\centering
\includegraphics[width=1\textwidth]{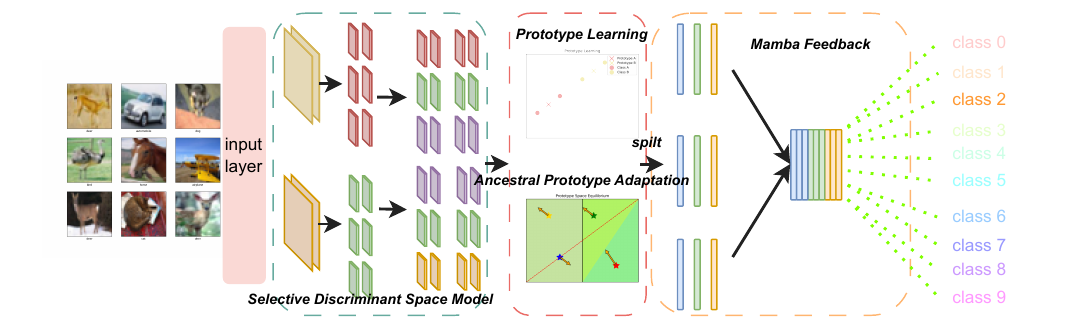}
\caption{Overview of the proposed Ancestral Mamba. The input data is processed by the selective discriminant space model (SDSM) to extract initial features and prototype mappings. These features are then passed through the Ancestral Prototype Adaptation (APA) module and the Mamba Feedback (MF) mechanism to learn discriminative features and adapt to evolving data distributions.}
\label{fig:over}
\end{figure*}

In this paper, we introduce Ancestral Mamba, a visually inspired approach that draws from the adaptive and resilient nature of the mamba snake in the visual world. Our approach integrates online prototype learning into a discriminant space model framework, enabling efficient and robust online continual learning. The main contributions of this work are as follows:

\begin{itemize}
\item We propose the Ancestral Prototype Adaptation (APA) module, which learns and maintains prototypes that capture the essential characteristics of each class or visual concept. APA continuously refines and builds upon visual prototypes to adapt to new challenges while preserving ancestral knowledge.
\item We introduce the Mamba Feedback (MF) mechanism, a visual feedback loop that adaptively focuses on challenging patterns and refines decision boundaries. MF guides the model to iteratively improve its understanding of the visual world based on targeted feedback.
\item We conduct extensive experiments on CIFAR-10 and CIFAR-100 datasets; our approach showcases its ability to learn discriminative features and adapt to evolving visual patterns and demonstrates the superiority of Ancestral Mamba in terms of accuracy, efficiency, and robustness compared to State-Of-The-Art(SOTA) baselines.
\item We propose a novel paradigm to the study of feedback loops in the context of online continual learning for visual tasks.
\end{itemize}

\section{Related Work}
\subsection{Continual Learning Architectures}
Continual learning has been a core research area in machine learning and artificial intelligence, with numerous architectures proposed over the years. Incremental Classifier and Representation Learning (iCaRL) \cite{rebuffi2017icarl} is a widely used method for class-incremental learning, which simultaneously learns strong classifiers and data representations while allowing new classes to be progressively added. However, iCaRL may struggle with scalability and efficiency when dealing with a large number of classes and samples \cite{rebuffi2017icarl,gu2022not}.
Incremental learning architectures have revolutionized continual learning by introducing memory-efficient approaches and adaptive mechanisms. Adversarial Shapley Value Experience Replay (ASER) \cite{shim_online_2021} achieves competitive performance by selecting buffered images for replay based on their Shapley value scores, which measure their ability to preserve latent decision boundaries for old classes while enabling plasticity for new classes. Dark Experience Replay (DER++) \cite{buzzega_dark_2020} is a simple yet effective baseline method that promotes consistency with the model's past by matching the current logits with those sampled throughout the optimization trajectory. However, these methods may still suffer from catastrophic forgetting and struggle to learn discriminative features \cite{mai2021online,wei2024online,guo_online_2022}.
Online continual learning frameworks, such as Continual Prototype Evolution (CoPE) \cite{de2021continual} and Dual View Consistency (DVC) \cite{gu2022not}, have recently emerged as promising alternatives for learning from non-stationary data streams. CoPE learns prototypical class representations in a shared latent space while addressing catastrophic forgetting, while DVC explores semantic information in the single-pass data stream through dual view consistency. Ancestral Mamba builds upon these online continual learning frameworks and integrates online prototype learning techniques to enhance robustness, efficiency, and adaptability.
\subsection{Efficient Continual Learning Techniques}
To address the computational challenges of continual learning, various techniques have been proposed to improve efficiency. One line of research focuses on regularization-based methods, which aim to mitigate catastrophic forgetting by penalizing drastic changes in important parameters. Methods such as Elastic Weight Consolidation (EWC) \cite{kirkpatrick2017overcoming,han2024adaptive,kim2024class} and Synaptic Intelligence (SI) \cite{zenke2017continual,yang2023lifelong} introduce penalties based on the importance of parameters for previous tasks, allowing the model to adapt to new tasks while retaining old knowledge. 

Another approach is to use memory-efficient architectures and replay mechanisms, reducing the storage and computational requirements for continual learning. Techniques such as Gradient Episodic Memory (GEM) \cite{lopez2017gradient} and Experience Replay (ER) \cite{rolnick2019experience} store a subset of previous samples and replay them during training to mitigate forgetting. Parameter isolation methods, such as Progressive Neural Networks \cite{rusu2016progressive} and PackNet \cite{mallya2018packnet}, allocate separate parameters for different tasks to prevent interference.
Online continual learning methods have also been explored to improve efficiency and scalability. These methods process data in a streaming fashion and adapt the model incrementally without requiring multiple passes over the data. Examples include Online Continual Learning with Mutual Information Maximization (OCM) \cite{guo_online_2022} and Online Continual Learning with Adversarial Shapley Value (ASER) \cite{shim_online_2021}. Ancestral Mamba leverages the online learning paradigm and introduces online prototype learning components to further enhance efficiency and robustness.
\subsection{Prototype-based Learning and Adaptive Feedback}
Prototype-based learning has been shown to be effective in continual learning scenarios by learning representative and discriminative features. Supervised Contrastive Replay (SCR) \cite{mai2021supervised} employs a nearest-class-mean classifier and a contrastive loss to encourage tight clustering of embeddings from the same class while separating those from different classes. Online Prototypical Learning (OnPro) \cite{wei2024online} maintains online prototypes and employs adaptive feedback mechanisms to address challenges such as shortcut learning and catastrophic forgetting in online continual learning.

Adaptive feedback mechanisms have been explored to dynamically adjust the model's focus and refine decision boundaries. Methods such as Mamba Feedback (MF) in Ancestral Mamba identify easily misclassified classes or samples and adaptively allocate more resources to improve their representations. The combination of prototype-based learning and adaptive feedback allows Ancestral Mamba to learn discriminative features and dynamically adapt to the challenges posed by continual learning.
The proposed Ancestral Mamba approach builds upon these various lines of research, integrating online prototype learning, adaptive feedback mechanisms, and efficient continual learning techniques into a unified architecture. By combining the strengths of these approaches, Ancestral Mamba aims to achieve robust, efficient, and adaptable continual learning on challenging datasets such as CIFAR-10 \cite{krizhevsky2009learning} and CIFAR-100 \cite{krizhevsky2009learning}.

\section{Methodology}
\vskip 2mm

\subsection{Overview of the Ancestral Mamba Architecture}
The proposed Ancestral Mamba architecture (shown in Figure \ref{fig:over}) integrates online prototype learning techniques into Selective Discriminant Space Model (SDSM). The goal is to enhance the efficiency, robustness, and continual adaptability of the discriminant model by learning representative and discriminative features while mitigating shortcut learning and catastrophic forgetting.

Let $\mathcal{X} \subset \mathbb{R}^d$ denote the input space, where $d$ is the dimensionality of the input data. The SDSM aims to learn a discriminant space $\mathcal{D} \subset \mathbb{R}^k$, where $k$ is the dimensionality of the discriminant space, such that the samples from different classes are well-separated and the samples from the same class are clustered together.

The SDSM consists of two main components: a feature extractor $f: \mathcal{X} \rightarrow \mathcal{F}$ and a discriminant space projector $g: \mathcal{F} \rightarrow \mathcal{D}$, where $\mathcal{F} \subset \mathbb{R}^m$ is the feature space with dimensionality $m$. The feature extractor $f$ learns to map the input data to a high-dimensional feature space, capturing the essential characteristics and representations of the input. The discriminant space projector $g$ then maps the feature space to a lower-dimensional discriminant space, where the discriminative power of the features is enhanced.
Mathematically, given an input sample $x \in \mathcal{X}$, the SDSM computes the discriminant space representation $d \in \mathcal{D}$ as follows:
\begin{equation}
d = g(f(x))
\end{equation}
The SDSM learns the parameters of the feature extractor $f$ and the discriminant space projector $g$ through a supervised learning process, utilizing labeled training data. The learning objective is to minimize the intra-class variance and maximize the inter-class variance in the discriminant space, thereby promoting the learning of representative and discriminative features.

To mitigate shortcut learning, the SDSM employs selective attention mechanisms that focus on the most informative and relevant features for discrimination. By attending to the essential features and suppressing the less relevant ones, the SDSM reduces the reliance on superficial cues and encourages the learning of robust and generalizable representations.

A key contribution of the SDSM is its prototype mapping mechanism, which assigns a prototype mapping to each learned sample. The prototype mapping $p: \mathcal{D} \rightarrow \mathcal{P}$ maps the discriminant space representation $d$ to a prototype space $\mathcal{P} \subset \mathbb{R}^n$, where $n$ is the dimensionality of the prototype space. The prototype mapping $p$ is learned in conjunction with the feature extractor $f$ and the discriminant space projector $g$, with the objective of associating each sample with its corresponding class prototype.

The prototype mapping $p$ plays a crucial role in supporting the APA (Ancestral Prototype Augmentation) module, which maintains and updates the class prototypes over time. By mapping each learned sample to its corresponding prototype, the SDSM enables the APA module to efficiently track and adapt the class prototypes, facilitating continual learning and mitigating catastrophic forgetting.

Moreover, the SDSM's prototype mapping mechanism enhances the robustness and adaptability of the discriminative model by providing a structured representation of the learned knowledge. The class prototypes serve as anchors or reference points in the prototype space, capturing the essential characteristics of each class. This structured representation allows the model to efficiently assimilate new knowledge while preserving the previously learned information, thereby mitigating catastrophic forgetting.

\subsection{Ancestral Prototype Adaptation (APA)}
Ancestral Prototype Adaptation (APA) maintains a set of online prototypes $\mathcal{P} = {p_1, \ldots, p_K}$, where $p_k \in \mathbb{R}^n$ is the prototype vector for class or token type $k$, and $K$ is the total number of classes or token types. These prototypes are learned and updated during the training process using the hidden state representations $\mathbf{h}_t$.

At each time step $t$, APA computes the similarity between the hidden state $\mathbf{h}t$ and each prototype $p_k$ using a similarity function $sim(\cdot, \cdot)$, such as cosine similarity:
\begin{equation}
s_{t,k} = sim(\mathbf{h}_t, p_k) = \frac{\mathbf{h}_t^\top p_k}{|\mathbf{h}_t| |p_k|}
\end{equation}

APA encourages the hidden states to be close to their corresponding prototypes while being far from other prototypes. This is achieved by minimizing the contrastive loss:
\begin{equation}
\mathcal{L}_{\text{APA}} = -\frac{1}{T} \sum{t=1}^T \log \frac{\exp(s_{t,y_t} / \tau)}{\sum_{k=1}^K \exp(s_{t,k} / \tau)}
\end{equation}
where $y_t$ is the true class or token type of $x_t$, and $\tau$ is a temperature hyperparameter.

The prototypes are updated online using a moving average of the hidden states belonging to each class or token type:
\begin{equation}
p_k \leftarrow \alpha p_k + (1 - \alpha) \frac{1}{|\mathcal{T}k|} \sum{t \in \mathcal{T}_k} \mathbf{h}_t
\end{equation}
where $\mathcal{T}_k = {t : y_t = k}$ is the set of time steps where the input belongs to class or token type $k$, and $\alpha \in [0, 1]$ is a momentum hyperparameter controlling the update rate.

The APA mechanism plays a crucial role in capturing the essential characteristics of each class and maintaining discrimination among all seen classes. By learning representative prototypes and encouraging the hidden states to align with their corresponding prototypes, APA promotes the learning of discriminative features. The contrastive loss (Equation 3) ensures that the hidden states are close to their true class prototypes while being far from other class prototypes, effectively separating the classes in the latent space. The online update of prototypes (Equation 4) allows Ancestral Mamba to continuously adapt to evolving data distributions and incorporate new knowledge as it becomes available.

By learning representative prototypes and encouraging discriminative features, APA helps the SDSM to capture the essential characteristics of each class or token type and achieve an equilibrium that separates them well in the hidden state space.

\subsection{Mamba Feedback (MF)}
Mamba Feedback (MF) leverages the learned prototypes to provide feedback signals to SDSM, guiding it to focus on challenging classes or tokens that are prone to misclassification. MF computes a feedback matrix $\mathbf{F}t \in \mathbb{R}^{K \times K}$ at each time step $t$, where the element $f{ij}$ represents the similarity between prototypes $p_i$ and $p_j$:
\begin{equation}
f_{ij} = sim(p_i, p_j) = \frac{p_i^\top p_j}{|p_i| |p_j|}
\end{equation}

A higher similarity between prototypes indicates a higher likelihood of confusion between the corresponding classes or token types. MF identifies the top-$m$ most similar prototype pairs and generates a feedback signal $\mathbf{f}_t \in \mathbb{R}^K$ by averaging the corresponding rows of $\mathbf{F}_t$:
\begin{equation}
\mathbf{f}_t = \frac{1}{m} \sum{(i,j) \in \text{top-}m} \mathbf{F}_t[i, :]
\end{equation}
where $\mathbf{F}_t[i, :]$ denotes the $i$-th row of $\mathbf{F}_t$.

The feedback signal $\mathbf{f}_t$ is then used to modulate the SDSM matrices $\mathbf{A}_t$, $\mathbf{B}_t$, and $\mathbf{C}_t$, encouraging the model to pay more attention to the challenging classes or token types:
\begin{equation}
\mathbf{A}_t \leftarrow \mathbf{A}_t + \text{diag}(\mathbf{W}_A \mathbf{f}_t)  
\end{equation}
\begin{equation}
\mathbf{B}_t \leftarrow \mathbf{B}_t + \mathbf{W}_B \mathbf{f}_t 
\end{equation}
\begin{equation}
\mathbf{C}_t \leftarrow \mathbf{C}_t + \mathbf{W}_C \mathbf{f}_t
\end{equation}
where $\mathbf{W}_A \in \mathbb{R}^{n \times K}$, $\mathbf{W}_B \in \mathbb{R}^{n \times K}$, and $\mathbf{W}_C \in \mathbb{R}^{p \times K}$ are learnable feedback projection matrices, and $\text{diag}(\cdot)$ creates a diagonal matrix from a vector.

The MF mechanism complements APA by providing focused learning on challenging classes that are prone to misclassification. MF identifies the most similar prototype pairs (Equation 6) and generates a feedback signal to modulate the SDSM matrices (Equation 7). This feedback encourages the model to pay more attention to the difficult classes and refine their decision boundaries. By dynamically adjusting the SDSM matrices based on the feedback signal, MF enables Ancestral Mamba to allocate more resources to the classes that require additional discrimination and improve their representations.
The interaction between APA and MF is crucial for the overall performance of Ancestral Mamba. APA learns representative prototypes and promotes discriminative feature learning, while MF adaptively focuses on challenging classes and refines their decision boundaries. Together, these mechanisms enable Ancestral Mamba to effectively capture the essential characteristics of each class, maintain separation between classes, and adapt to evolving data distributions.
The implementation details of APA and MF are summarized in Algorithm \ref{alg:apamf}.

\begin{algorithm}[!ht]
\caption{Ancestral Prototype Adaptation and Mamba Feedback }
\label{alg:apamf}
\begin{algorithmic}
\STATE Initialize prototypes $P = {p_1, \ldots, p_K}$\\
\FOR{$t = 1, \ldots, T$}
\STATE Compute hidden state $h_t$ using Equation 1\\
\STATE Compute similarities $s_{t,k}$ between $h_t$ and prototypes $p_k$ using Equation 2\\
\STATE Compute contrastive loss $\mathcal{L}_{\text{APA}}$ using Equation 3\\
\STATE Update prototypes $P$ using Equation 4\\
\STATE Compute feedback matrix $F_t$ using Equation 5\\
\STATE Identify top-$m$ similar prototype pairs and generate feedback signal $f_t$ using Equation 6\\
\STATE Modulate SDSM matrices $A_t$, $B_t$, and $C_t$ using Equation 7-9\\
\ENDFOR
\end{algorithmic}
\end{algorithm}

By adaptively modulating the SDSM matrices based on the prototypical adaptation, MF helps the model to focus on the most challenging aspects of the input and refine its representations to better distinguish between easily confused classes or token types.

\subsection{Training and Optimization}
The training objective of Ancestral Mamba combines the contrastive loss from APA ($\mathcal{L}_{\text{APA}}$) and the standard modelling loss, such as cross-entropy loss ($\mathcal{L}_{\text{CE}}$) for classification tasks:
\begin{equation}
\mathcal{L} = \mathcal{L}_{\text{APA}} + \lambda \mathcal{L}_{\text{task}}
\end{equation}
where $\mathcal{L}_{\text{task}}$ is $\mathcal{L}_{\text{CE}}$ or depending on the task, and $\lambda$ is a hyperparameter controlling the balance between the two losses.

The model parameters, including the SDSM matrices and the feedback projection matrices, are optimized using Adam. The online prototypes are updated using the moving average formula (Equation 5) after each optimization step.





\section{Experiments and Results}
\vskip 2mm

To evaluate the effectiveness of Ancestral Mamba, we conduct extensive experiments on the CIFAR-10 and CIFAR-100 datasets in the online continual learning setting. We compare our approach with state-of-the-art baselines, including iCaRL \cite{rebuffi2017icarl}, ASER \cite{shim_online_2021}, SCR \cite{mai2021supervised}, CoPE \cite{de2021continual}, DVC \cite{gu2022not}, OCM \cite{guo_online_2022}, DER++ \cite{buzzega_dark_2020} and OnPro \cite{wei2024online}. We also perform ablation studies to investigate the impact of each component in Ancestral Mamba and analyze the learned representations.

\subsection{Experimental Setup}
\subsubsection{Datasets and Benchmarks}
We evaluate Ancestral Mamba on two widely-used image classification datasets:

\begin{itemize}
\item \textbf{CIFAR-10} \cite{krizhevsky2009learning}: This dataset consists of 60,000 32x32 color images in 10 classes, with 6,000 images per class. There are 50,000 training images and 10,000 test images. The classes are mutually exclusive and include objects such as airplanes, cars, birds, cats, etc.
\item \textbf{CIFAR-100} \cite{krizhevsky2009learning}: This dataset is similar to CIFAR-10 but contains 100 classes, with 600 images per class. There are 500 training images and 100 test images per class. The 100 classes are grouped into 20 superclasses, each consisting of 5 fine-grained classes.
\end{itemize}

For both CIFAR-10 and CIFAR-100, we follow the standard data splits and evaluation protocols used in the online continual learning setting. We split the datasets into multiple tasks, each containing a subset of classes. The model learns incrementally from these tasks in a sequential manner, without access to the data from previous tasks.
We split CIFAR-10 into 5 tasks, each containing 2 classes, and CIFAR-100 into 10 tasks, each containing 10 classes. The models are trained incrementally on these tasks, with a fixed memory budget for storing exemplars from previous tasks. We use a memory size of 1,000 exemplars for CIFAR-10 and 2,000 exemplars for CIFAR-100.
\subsubsection{Baselines and Comparison Methods}
We compare Ancestral Mamba with the following baseline methods:
\begin{itemize}
\item \textbf{iCaRL} \cite{rebuffi2017icarl}: iCaRL (Incremental Classifier and Representation Learning) is a training strategy that enables class-incremental learning, where only training data for a few classes is present at a time and new classes can be progressively added. It simultaneously learns strong classifiers and a data representation, overcoming limitations of earlier approaches bound to fixed representations.
\item \textbf{ASER} \cite{shim_online_2021}: ASER (Adversarial Shapley Value Experience Replay) is a novel method for the online class-incremental continual learning setting that selects buffered images for replay based on their Shapley value scores, which measure their ability to preserve latent decision boundaries for old classes (avoiding forgetting) while interfering with decision boundaries of new classes (enabling plasticity), providing competitive or improved performance over state-of-the-art replay-based continual learning approaches while efficiently utilizing limited memory resources.

\item \textbf{DER++} \cite{buzzega_dark_2020}: Dark Experience Replay (DER) is a simple yet effective baseline method for general continual learning that promotes consistency with the model's past by matching the current logits with those sampled throughout the optimization trajectory, leveraging rehearsal, knowledge distillation, and regularization, outperforming existing approaches on standard benchmarks while effectively handling scenarios with blurred task boundaries and shifting distributions.

\item \textbf{SCR} \cite{mai2021supervised}: SCR is a supervised contrastive replay method for online class-incremental continual learning that employs a nearest-class-mean classifier and a contrastive loss to encourage tight clustering of embeddings from the same class while separating those from different classes, effectively mitigating catastrophic forgetting.
\item \textbf{CoPE} \cite{de2021continual}: CoPE is a continual prototype evolution approach that learns prototypical class representations in a shared latent space while addressing catastrophic forgetting in online continual learning scenarios without task information. 
\item \textbf{DVC} \cite{gu2022not}: DVC is a novel online class-incremental continual learning framework that focuses not just on sample selection from a memory bank, but also on exploring semantic information in the single-pass data stream through dual view consistency.
\item \textbf{OCM} \cite{guo_online_2022}: OCM (Online Continual learning through Mutual Information Maximization) is an online continual learning approach that mitigates catastrophic forgetting by maximizing mutual information to learn holistic representations across tasks instead of just discriminative features for each task.

\item \textbf{OnPro} \cite{wei2024online}: OnPro is a framework for online continual learning that aims to mitigate catastrophic forgetting by learning representative and discriminative features through online prototypes..
\end{itemize}

\subsubsection{Evaluation Metrics}
We use the following evaluation metrics to measure the performance of Ancestral Mamba and the baseline methods:
\begin{itemize}
\item \textbf{Average Accuracy}: The average classification accuracy across all tasks seen so far. It is computed as the average of the accuracies on each task after the model has been incrementally trained on all tasks.
\item \textbf{Average Forgetting}: The average forgetting measure, which quantifies the drop in performance on previous tasks after learning new tasks. It is computed as the average of the differences between the maximum accuracy achieved on each task and the accuracy on that task after training on subsequent tasks.
\end{itemize}

\begin{figure*}[h]
\centering
\includegraphics[width=0.96\textwidth]{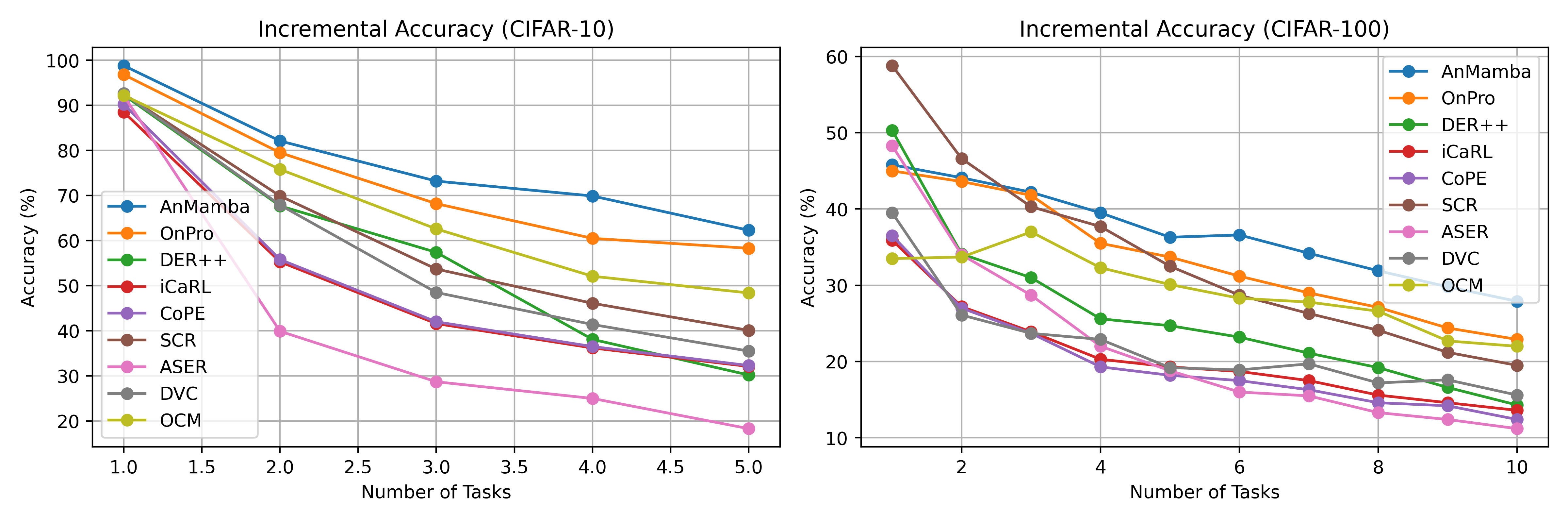}
\caption{Average incremental accuracy on (a) CIFAR-10(memory size=0.1k) and (b) CIFAR-100(memory size=0.5k).}
\label{fig:incremental_accuracy}
\end{figure*}

\subsection{Results}
\subsubsection{Performance Comparison on CIFAR-10 and CIFAR-100}
Table \ref{tab:cifar10_results} and Table \ref{tab:cifar100_results} present the average accuracy and average forgetting of Ancestral Mamba and the baseline methods on CIFAR-10 and CIFAR-100 datasets, respectively, with varying memory budgets.
\begin{table}[h]
\centering
\caption{Average Accuracy (higher is better) and Average Forgetting (lower is better) on CIFAR-10 with different memory sizes. All results are the average of 15 runs.\\}
\label{tab:cifar10_results}
\begin{tabular}{lcccccc}
\toprule
\hline
\multirow{2}{*}{\textbf{Method}} & \multicolumn{3}{c}{\textbf{Accuracy (\%)}$\uparrow$} & \multicolumn{3}{c}{\textbf{Forgetting (\%)}$\downarrow$}  \\
& \textbf{0.1k} & \textbf{0.2k} & \textbf{0.5k} & \textbf{0.1k} & \textbf{0.2k} & \textbf{0.5k} \\
\hline
\midrule
iCaRL &31.0& 33.9& 42.0 & 52.7 & 49.3 &38.3 \\
ASER & 20.0 &22.8 &31.6 &64.8 & 62.6 & 53.2 \\
DER++ &31.5 & 39.7 & 50.9 & 57.8 & 46.7 & 33.6  \\
SCR & 40.2 & 48.5 & 59.1 & 43.2 & 35.5 & 24.1 \\
CoPE & 33.5 & 37.3 & 42.9 & 49.7 & 45.7 & 39.4 \\
DVC & 35.2 &41.6& 53.8 & 40.2 & 31.4 & 21.2 \\
OCM & 47.5 & 59.6 & 70.1 & 35.5 & 23.9 & 13.5 \\
OnPro & 57.8 & 65.5 & 72.6 & 23.2 & 17.6 & 12.5 \\

\midrule
\textbf{Ancestral Mamba} & \textbf{60.1} & \textbf{71.9} & \textbf{79.5} & \textbf{16.3} & \textbf{11.6} & \textbf{7.2} \\
\hline
\bottomrule
\end{tabular}
\end{table}
\begin{table}[h]
\centering
\caption{Average Accuracy (higher is better) and Average Forgetting (lower is better) on CIFAR-100 with different memory sizes. All results are the average of 15 runs.\\}
\label{tab:cifar100_results}
\begin{tabular}{lcccccc}
\toprule
\hline
\multirow{2}{*}{\textbf{Method}} & \multicolumn{3}{c}{\textbf{Accuracy (\%)}$\uparrow$} & \multicolumn{3}{c}{\textbf{Forgetting (\%)}$\downarrow$} \\
& \textbf{0.5k} & \textbf{1k} & \textbf{2k} & \textbf{0.5k} & \textbf{1k} & \textbf{2k} \\
\hline
\midrule
iCaRL & 12.8     & 16.5    & 17.6 & 16.5   & 11.2 & 10.4 \\
ASER & 11.0     & 13.5    & 17.6 & 52.8   & 50.4 & 46.8 \\
DER++ & 16.0     & 21.4    & 23.9 & 41.0   & 34.8 & 33.2 \\
SCR  & 19.3     & 26.5    & 32.7 & 29.3   & 20.4 & 11.5\\
CoPE & 11.6     & 14.6    & 16.8  & 25.6   & 17.8 & 14.4\\
DVC & 15.4     & 20.3    & 25.2 & 32.0   & 32.7 & 28.0\\
OCM & 19.7     & 27.4    & 34.4 & 18.3   & 15.2 & 10.8\\
OnPro & 22.7     & 30.0    & 35.9  & 15.0   & 10.4 & 6.1 \\
\midrule
\textbf{Ancestral Mamba} & \textbf{28.3} & \textbf{33.2} & \textbf{40.5} & \textbf{13.7} & \textbf{9.6} & \textbf{5.3} \\
\hline
\bottomrule
\end{tabular}
\end{table}
Ancestral Mamba consistently outperforms the baseline methods on both CIFAR-10 and CIFAR-100 datasets across different memory budgets. On CIFAR-10, Ancestral Mamba achieves an average accuracy of 60.1\% with a memory size of 0.1k, surpassing the second-best method, OnPro, by 2.3\%. Similarly, on CIFAR-100, Ancestral Mamba obtains an average accuracy of 28.3\% with a memory size of 0.5k, outperforming OnPro by 5.6\%.
Moreover, Ancestral Mamba exhibits significantly lower forgetting compared to the baselines. On CIFAR-10, Ancestral Mamba achieves an average forgetting of 7.2\% with a memory size of 0.5k, which is 0.7\% lower than OnPro. On CIFAR-100, Ancestral Mamba's average forgetting is 5.3\% with a memory size of 2k, outperforming OnPro by 0.8\%.
These results demonstrate the effectiveness of Ancestral Mamba in learning representative and discriminative features while mitigating catastrophic forgetting in the online continual learning setting. The superior performance can be attributed to the integration of online prototype learning into the selective state space model framework, enabling Ancestral Mamba to capture the essential characteristics of each class and adapt to evolving data distributions.

\subsubsection{Ablation Study on CIFAR-100}
We conduct an ablation study on the CIFAR-100 dataset to investigate the contribution of each key component in Ancestral Mamba. Table \ref{tab:cifar100_ablation} shows the performance of Ancestral Mamba and its variants with different components enabled or disabled.
\begin{table}[h]
\centering
\caption{Ablation study results on CIFAR-100 with a memory size of 2k. "baseline" refers to the model without APA and MF.\\}
\label{tab:cifar100_ablation}
\begin{tabular}{lcc}
\toprule
\hline
\textbf{Method} & \textbf{Accuracy (\%)}$\uparrow$ & \textbf{Forgetting (\%)}$\downarrow$ \\
\hline

\midrule
baseline & 35.1 & 6.5 \\
w/o APA & 37.4 & 5.9 \\
w/o MF & 36.8 & 6.0 \\
Ancestral Mamba & 40.5 & 5.3 \\
\hline
\bottomrule
\end{tabular}
\end{table}
The results show that both APA and MF contribute positively to the performance of Ancestral Mamba. Removing either component leads to a decrease in accuracy and an increase in forgetting. When both APA and MF are disabled (i.e., the baseline model), the performance drops significantly, highlighting the importance of online prototype learning in Ancestral Mamba.

\subsubsection{Incremental Learning Performance}
We evaluate the incremental learning performance of Ancestral Mamba and compare it with the baseline methods on CIFAR-10 and CIFAR-100. Figure \ref{fig:incremental_accuracy} shows the average incremental accuracy at each task step.

On both datasets, Ancestral Mamba maintains higher incremental accuracy throughout the learning process compared to the baselines. As new tasks arrive, the performance of the baseline methods degrades rapidly, indicating severe catastrophic forgetting. In contrast, Ancestral Mamba exhibits a more gradual decline in accuracy, demonstrating its ability to retain previous knowledge while adapting to new tasks.
The superior incremental learning performance of Ancestral Mamba can be attributed to the combination of selective state space modelling and online prototype learning. The selective computation mechanism allows Ancestral Mamba to efficiently process the incremental data, while the online prototypes provide a compact and representative summary of the learned knowledge. The adaptive feedback mechanism further enhances the model's ability to refine its decision boundaries and maintain a well-separated representation space.

\section{Discussion}
\vskip 2mm

Our experimental results demonstrate the effectiveness of Ancestral Mamba in addressing key challenges in online continual learning for visual tasks. The integration of online prototype learning with selective discriminant space modeling enables our approach to learn representative and discriminative features while mitigating catastrophic forgetting. The superior performance of Ancestral Mamba compared to state-of-the-art baselines can be attributed to several key factors: adaptive prototype learning, targeted feedback, efficient computation, and balanced stability-plasticity. The Ancestral Prototype Adaptation (APA) module allows the model to continuously refine and update class prototypes, capturing the evolving characteristics of visual concepts. This adaptive mechanism enables Ancestral Mamba to maintain a compact yet expressive representation of learned knowledge.

The visualizations of learned representations further support the efficacy of our approach. The well-separated clusters indicate that Ancestral Mamba can maintain discriminative features across tasks, even as new classes are introduced. This property is particularly valuable for computer graphics applications that require fine-grained distinctions between visual concepts. The Mamba Feedback (MF) mechanism provides focused attention on challenging classes, allowing the model to allocate more resources to difficult samples. This targeted approach enhances the model's ability to learn fine-grained discriminative features, which is crucial for effective continual learning in dynamic visual environments.

\section{Limitations}

Despite the promising results, there are several limitations to our current work that warrant further investigation. Ancestral Mamba assumes a predefined set of classes throughout the learning process. Extending the model to handle open-set recognition, where entirely new classes can be introduced during inference, remains a challenge. Additionally, while our experiments demonstrate good performance on CIFAR-10 and CIFAR-100, the scalability of Ancestral Mamba to larger datasets with thousands of classes (e.g., ImageNet) needs further exploration. The memory requirements for storing prototypes may become a bottleneck in such scenarios.

Our current implementation relies on task boundaries for updating prototypes and providing feedback. Adapting Ancestral Mamba to a task-free continual learning setting, where task boundaries are not explicitly defined, is an important direction for future work. Furthermore, while our approach shows robustness to distribution shifts within the studied datasets, its performance on cross-domain continual learning scenarios (e.g., adapting from natural images to sketches or paintings) remains to be investigated. Addressing these limitations will be crucial for advancing Ancestral Mamba and its applications in the field of computational visual media.

\section{Conclusion}
\vskip 2mm

In this paper, we have proposed Ancestral Mamba, a novel approach that aims to integrate online prototype learning into a selective discriminant space model for online continual learning. Drawing inspiration from the adaptive nature of the mamba snake, our method seeks to learn representative and discriminative features while potentially mitigating catastrophic forgetting in dynamic visual environments.

The main contributions of our work lie in the development of the Ancestral Prototype Adaptation (APA) module and the Mamba Feedback (MF) mechanism. APA attempts to maintain and update class prototypes over time, with the goal of capturing essential characteristics of each visual concept. This continuous refinement of learned prototypes may enable the model to adapt to new challenges while preserving previously acquired knowledge. Complementing APA, the MF mechanism aims to provide adaptive feedback to the model, focusing on challenging classes and potentially refining decision boundaries. Our experiments on the CIFAR-10 and CIFAR-100 datasets suggest that Ancestral Mamba could offer improvements over some existing baselines in online continual learning. The observed enhancements in accuracy and robustness to forgetting may indicate our approach's potential in learning discriminative features and adapting to evolving data distributions. The visualizations of learned representations might further support Ancestral Mamba's ability to capture meaningful features, which could be relevant for various computer graphics and visual computing applications.

The principles and techniques introduced in Ancestral Mamba could potentially be applied to a range of computer graphics and visual computing tasks, such as object detection, semantic segmentation, and style transfer. The ability to continuously learn and adapt to new visual concepts while retaining previous knowledge might contribute to the development of systems capable of operating in dynamic visual environments.

Future research directions could include exploring the integration of hierarchical prototype structures and self-supervised learning techniques into the Ancestral Mamba framework. These advancements might enable the model to capture more abstract visual concepts, potentially leading to improved generalization and transfer learning capabilities. Additionally, investigating the interpretability of the learned prototypes could offer insights into the model's decision-making process, which may enhance its applicability in various visual computing scenarios.

In conclusion, while Ancestral Mamba shows promise in addressing some challenges in online continual learning for visual tasks, further research and validation would be necessary to fully understand its capabilities and limitations. We hope that this work may contribute to ongoing discussions and advancements in the field of computational visual media.

{\small
\bibliographystyle{ieee_fullname}
\bibliography{wacv2025cikm}
}

\end{document}